\documentclass[letter,usenatbib]{aa}

\usepackage{amssymb}
\usepackage{amsmath}
\usepackage{graphicx}
\usepackage{natbib}
\usepackage{hyperref}
\usepackage{pdflscape}
\usepackage{longtable}
\usepackage{multicol}
\usepackage{multirow}
\usepackage{url}
\usepackage{graphicx}
\usepackage{color}
\usepackage{txfonts}

\DeclareRobustCommand{\ION}[2]{%
\relax\ifmmode
\ifx\testbx\f@series
{\mathbf{#1\,\mathsc{#2}}}\else
{\mathrm{#1\,\mathsc{#2}}}\fi
\else\textup{#1\,{\mdseries\textsc{#2}}}%
\fi}

\newcommand{\hii}{\ION{H}{ii}}

\newcommand{\alf}{[$\alpha$/Fe]}

\newcommand{\age}{$\mathcal{A}_{\star,L}$}
\newcommand{\met}{$\mathcal{Z}_{\star,L}$}

\usepackage{txfonts}
\usepackage[]{hyperref}
\begin{document}

   \title{\alf\ traced by \ION{H}{ii} regions from the CALIFA survey:}
   
   \subtitle{The connection between morphology and chemical abundance patterns}

   \author{ S.~F.~S\'anchez\inst{\ref{unam}}
   \and C.~Espinosa-Ponce\inst{\ref{unam}}
   \and L.~Carigi\inst{\ref{unam}}
   \and C.~Morisset\inst{\ref{unam2}}
   \and J.~K.~Barrera-Ballesteros\inst{\ref{unam}}\\
   \and C.~J.~Walcher\inst{\ref{aip}}
   \and R.~Garc\'\i a-Benito\inst{\ref{iaa}}
   \and A.~Camps-Fari\~na\inst{\ref{unam}}
   \and L. Galbany\inst{\ref{gran}}
          }
\institute{Instituto de Astronom\'ia, Universidad Nacional Auton\'oma de M\'exico, A.P. 70-264, 04510, M\'exico, CDMX \label{unam}
\and Universidad Nacional Autónoma de México, Instituto de Astronomía, AP 106,  Ensenada 22800, BC, M\'exico \label{unam2}
\and Leibniz-Institut für Astrophysik Potsdam (AIP), An der Sternwarte 16, 14482 Potsdam, Germany \label{aip}
\and Instituto de Astrof\'isica de Andaluc\'ia (IAA/CSIC), Glorieta de la Astronom\'{\i}a s/n Aptdo. 3004, E-18080 Granada, Spain,\label{iaa} 
\and Departamento de F\'isica Te\'orica y del Cosmos, Universidad de Granada, E-18071 Granada, Spain.\label{gran}
}

   \date{Received ---, 2021; accepted ---, 29021}

 
  \abstract
   {Differential enrichment between $\alpha$- and Fe-peak elements is known to be strongly connected with the shape of the star formation history (SFH), the star formation efficiency (SFE), the inflow and outflow of material, and even the shape of the Initial Mass Function (IMF). However, beyond the Local Group detailed explorations are mostly limited to early-type galaxies due to the lack of a good proxy for \alf\ in late-type ones, limiting our understanding of the chemical enrichment process.}
   {We intent to extend the explorations of \alf\ to late-type galaxies, in order to understand the details of the differential enrichment process.}
   {We compare the gas phase oxygen abundance with the luminosity weighted stellar metallicity in an extensive catalog of $\sim$25,000 \hii\ regions extracted from the Calar Alto Legacy Integral Field Area (CALIFA) survey, an exploration using integral-field spectroscopy of $\sim$900 galaxies, covering a wide range of masses and morphologies. This way we define [O/Fe] as the ratio between both parameters, proposing it as an indirect proxy of the \alf\ ratio.{ This procedure is completely different from the one adopted to estimate \alf\ from high resolution spectroscopic data for stars in our Galaxy.}}
   {We illustrate how the [O/Fe] parameter describes the chemical enrichment process in spiral galaxies, finding that: (i) it follows the decreasing pattern with [Fe/H] reported for the \alf\ ratio and (ii) its absolute scale depends of the stellar mass and the morphology. We reproduce both patterns using two different chemical evolution models (ChEM), considering that galaxies with different stellar mass and morphology present (i) different SFHs, SFEs and different inflow/outflow rates, or (ii) a different maximum stellar mass cut for the IMF. We will explore the differential chemical enrichment using this new proxy galaxy by galaxy and region by region in further studies.}
   {}

   \maketitle
%
\section{Introduction}
\label{sec:intro}

All metals in the Universe are produced by the thermonuclear fusion reactions that are the core engine of stars. For intermediate- and low-mass stars, a fraction of these metals are expelled into the interstellar medium (ISM) along their life-time as part of the stellar winds and, in particular in their later phases, when stars lose their envelops. However, most of those metals remain inside the stars and end up in white dwarfs, neutron stars and black holes \citep[e.g.][]{koba20}. The metals that enrich the ISM, traced by the most abundant elements, are produced during supernovae explosions. 
In particular $\alpha-$elements { \citep[O, Mg, Si, S, Ar, Ca and Ti, e.g.,][]{matt92}}, produced mostly in massive, short-lived stars, are transferred to the ISM when these stars explode as core-collapse supernovae, either Type-II or Type-Ib and Ic { \citep[e.g.][]{woos95})}. Therefore, the production and enrichment of { these heavy metals} is directly associated with the star formation process. On the other hand, iron peak elements are produced in stars of a wide range of masses, and their ISM enrichment is dominated by Type Ia supernovae (SNIa), which are triggered in binary systems { \citep{koba20}}. These events are not, in principle, connected with the most recent star formation processes \citep[e.g.][]{galbany:2014}, as they happen { during the life-time} of a stellar population after a certain delay time \citep[e.g.][]{walcher16,cast21}, and span a wide time interval.

{ The most detailed explorations of the differential abundance between these two families of elements and its connection with the evolution of the stellar populations comes from the analysis of high- and intermediate-resolution optical and NIR spectra of hundreds to hundred of thousand stars at different locations within our Galaxy \citep[e.g.][]{apogee,lamost,galah,gaia}. Those analysis rely on the comparison of the full observed spectra, or a particular set of absorption features, with predictions from theoretical stellar atmosphere models \citep[e.g.][]{apogee} or previously well labeled empirical observations \citep[e.g.][]{galah}. Among their several results they found that (i) disk stars, in particular those in the inner disk, follow a well defined track between \alf\ and  [Fe/H], with a slope near $\sim-$0.3; (ii) stars above the thin disk present a clear super-solar \alf\ enhancement; and (iii) this enhancement is directly connected with the age of the stellar population \citep[e.g.][ Fig.8]{lamost}. These results agree with the idea that the different time-scales for the production of $\alpha$ and iron-peak elements can be used as a clock for the star-formation history \citep[SFH,][]{matt92,matt99,thomas05}.}

{ Beyond our galaxy, the derivation of the \alf\ ratio relies on the
comparison of the strength of certain absorption features or the full observed spectra for unresolved stellar populations with the predictions by stellar synthesis codes
\citep[e.g.][]{trager+00,walcher09,vazdekis15}.} For this reason, most of these explorations are limited to study elliptical galaxies, whose spectra present strong metal absorption features, and a more limited contamination by ionized gas emission lines \citep[e.g.][]{conroy12,walcher15}. { Only until very recently this kind of exploration has been extended to late-type galaxies too \citep[][]{watson21}. These studies have found that (i) early-type galaxies exhibit the same negative trend between \alf\ and [Fe/H] reported for the resolved stellar populations in our galaxy \citep{walcher16}, and (ii) the \alf\ ratio decreases with morphology \citep{watson21}. We should note that in non of those cases the galaxies were explored as resolved entities.}

In this study we propose a different approach, using the presence of strong ionized gas emission lines to our advantage. Instead of tracing \alf\ using the information provided by the absorption features in the stellar population, we gauge the $\alpha$-element abundance traced by the oxygen abundance (O/H) in the ISM. Ever since the pioneering studies by \citet{sear71,comte74} and \citet{peim78}, \hii\ regions have been used to trace the chemical content in spiral galaxies { \citep[e.g.][]{sanchez14}, including the Milky-Way \citep[e.g.][]{esteban18}}. These gaseous clouds trace the current metal content of the interstellar medium, assumed to be the same as that of the short-lived OB-stars that ionized them (i.e., the most recent generation
of stars). However, this metal content is the consequence of the full chemical enrichment history (ChEH) at the location in which the
star formation is triggered and, therefore, the HII created \citep{ARAA}. 
{ Due to that it is possible to combine the} oxygen abundance estimates derived from the analysis of the emission lines observed in the optical spectra of \hii\ regions, with the iron abundance estimated from the analysis of the underlying stellar spectrum { to estimate of the \alf\ ratio, despite the fact that both measurements are tracing the metal content of two different populations. We demonstrate that this} is a useful approach to explore this elusive property in spiral galaxies{, that we will exploit in detailed in a forthcoming article (Espinosa-Ponce et al. in prep.)}.
Previous attempts in the same direction \citep[e.g.][]{lian18,ARAA}, compared the two parameters separately, reaching similar conclusions. { This paper is organized as follows.  In Sect. \ref{sec:data}  we describe our data, presenting the properties of the sample of galaxies and \hii\ regions and a summary of the analysis performed on the data to derive the explored physical parameters; in Sec. \ref{sec:res} we present the main results of our analysis, showing the observational trends and compare them with the expectations obtained by means of chemical evolution models. Finally, we summarize our main conclusions in Sec. \ref{sec:con}.}


\section{Data and Analysis}
\label{sec:data}

We use the catalog of properties of \hii\ regions and aggregations published by \citet{espi20}\footnote{\url{http://ifs.astroscu.unam.mx/CALIFA/HII\_regions/}}. These \hii\ regions, and their { spectroscopic} properties, were extracted from the data provided by the Calar Alto Legacy Integral Field Area survey \citep[CALIFA][]{califa}. { This survey explored the full optical extension of a sample of galaxies representative of the bulk population in the nearby universe ($<$100 Mpc) \citep{walcher14}, using the PPAK Integral Field Unit \citep[][]{kelz06}. For further details on the sample, survey strategy, observations and reduction consult \citet{walcher14}, \citet{sanchez16} and \citet{pisco}}
   \begin{figure*}
   \centering
   \resizebox{0.99\hsize}{!}{\includegraphics{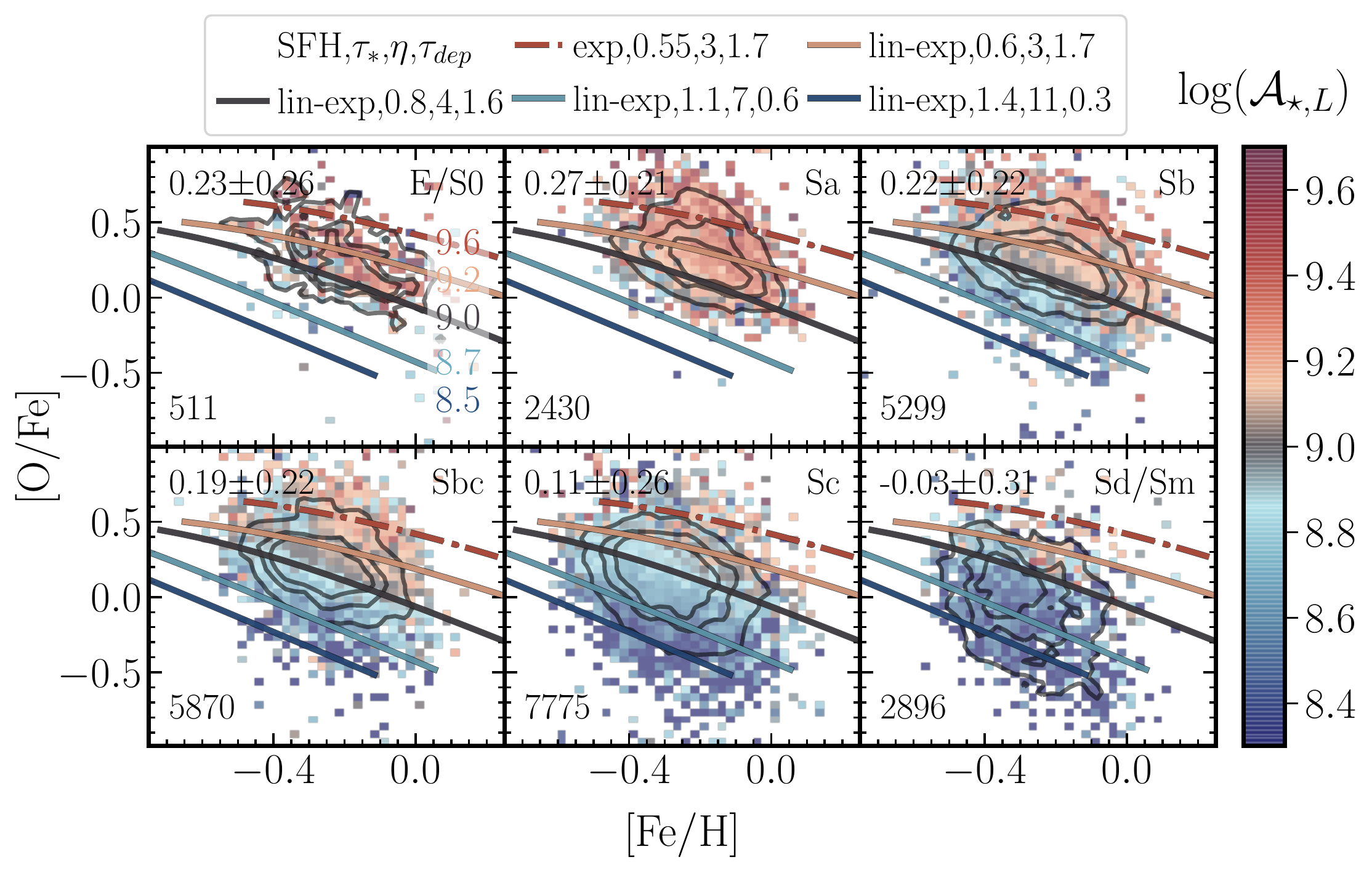}}
   \caption{{ Distribution of [O/Fe] as a function of [Fe/H], for the \hii\ regions of our sample (segregated by { galaxy} morphology), color coded by the average \age\ of the underlying stellar population. Contours represent the density of the plotted regions {, encircling 95\%,80\% and 40\% of them, respectively,} with the total number shown at { the bottom-left of} each panel. { The average [O/Fe] value for each morphology is shown at the top-left of each panel too}.
   { The solid and dashed-dotted lines are all the same in the six panels. They correspond to the predictions by the five ChEMs described in the text, generated to cover the observed distributions}. Each model is defined by the values of a set of input parameters ($\tau_*$, $\eta$, $\tau_{dep}$) and two shapes of the SFHs (exponential decay, $exp$, and linear rising plus exponential decay, $lin-exp$), indicated in the upper legend.
   Lines are color coded according to the average \age\ too. { Note that there is no exact one-to-one correspondence between a single model and the observed distribution for a single morphological bin (i.e., panel). }} \label{fig:OFe}}%
    \end{figure*}

The emission lines and stellar population properties are derived using the broadly tested {\tt Pipe3D} pipeline \citep[e.g.][]{pipe3d_ii,ibarra19,sanchez20}. This pipeline performs a spectral fitting decomposition of the stellar continuum based on a combination of SSPs, convolved and shifted using a Gaussian kernel to account for the line-of-sight velocity distribution, and attenuated by an extinction law. 
Once the best stellar model is obtained for each spectrum in each cube, this model is subtracted from the original spectra creating a cube with only the ionized emission line component. 
{ Then, this cube is then used}
to derive the main properties of a set of pre-defined emission lines. 

\citet{espi20} extracted a catalog of \hii\ regions and aggregations from the dataproducts provided by {\sc Pipe3D} applied to the CALIFA datacubes. In order to do so they followed the prescriptions by \citet{sanchez12b}, selecting clumpy structures in the H$\alpha$ intensity map of each galaxy, and selecting from them those compatible with being ionized by a young stellar population: i.e., with a fraction of young stars and an EW(H$\alpha$) high enough to produce the observed ionization. 
This analysis provided a catalog that comprises the main properties of the ionized gas emission lines (including the flux intensities of 52 lines) and the main physical parameters of the underlying stellar population, for a total of $\sim$25,000 \hii\ regions across the 924 explored galaxies.


We make use of this catalog, extracting three main physical properties: ($i$) the luminosity weighted age (\age) of the underlying stellar population, that corresponds to the first moment of the Age Distribution Function in light (ADF), directly provided by the stellar decomposition performed by {\sc Pipe3D}.
This parameter is directly connected with how sharp the star formation history is in galaxies,{ with older stellar populations having SFHs which peak at earlier epochs}, \citep[e.g.][]{rgb17}; ($ii$) the luminosity weighted metallicity normalized to the solar value { [Z/H] or \met)}. This is the first moment of the Metallicity Distribution Function (MDF) in light at the same band { as the ADF }\citep[e.g.]{mejia20}.
Due to the nature of the adopted SSPs, this parameter is a direct proxy { of } the luminosity-weighted iron abundance, [Fe/H], for a solar abundance of 12+log(Fe/H)=7.50 \citep{aspl09}. It is worth noting that we derive [Z/H] using SSP-models with solar composition (i.e., \alf=0). This may create a bias in [Fe/H] towards higher values, that can be as large as $\sim$30\%. However, this bias does not affect either our qualitative results or the reported trends, although it may produce relative offsets in the absolute reported abundances. Hereafter we refer to this parameter as [Fe/H]; and ($iii$) the gas phase oxygen abundance, 12+log(O/H), derived using the calibrator presented by \citet{ho19}. This calibrator uses a set of emission line ratios, and associates them with oxygen abundances measured using the direct method { applying} a neural-network procedure. We should note that the use of this or other of the calibrators applied to derive the oxygen abundance for the considered catalog do not alter qualitatively the results presented here (Espinosa-Ponce et al., in prep.). We transform this value to [O/H] by adopting a solar oxygen abundance of 12+log(O/H)$_\odot$=8.69 \citep{aspl09}.




\section{Results}
\label{sec:res}

Figure \ref{fig:OFe} shows the distribution of the relative oxygen-to-iron abundance, [O/Fe], as a function of [Fe/H] for the catalog of \hii\ regions explored along this article. [O/Fe] is constructed from the [O/H] and [Fe/H] parameters described in the previous section. 
We should recall that while [Fe/H] is an average of the iron abundances of the surviving stars, [O/H] traces the oxygen content of the most recently formed ones \citep[e.g.,][]{rosa14b}. 
However, [O/H] is also a consequence of the evolution of the stellar population, clearly connected with its age \citep[e.g., ][]{sanchez15,espi20}, and more strongly with the stellar mass at any look-back time \citep[e.g., Fig. 1][]{maio19}. { Furthermore, a fraction of the oxygen is depleted into dust in \hii\ regions, and therefore the [O/H] (and [O/Fe]) reported here may be $\sim$0.08-0.12 dex lower than if they were both measured in stars \citep{peim10}. Thus,} the [O/Fe] shown in here should not be interpreted as the $\alpha$-enhancement of either the average population (luminosity or mass weighted) at the location of the \hii\ region, or of a particular subset of the population. However, as we will demonstrate later, it still retains valuable information about the $\alpha$-enhancement process along the cosmological evolution of the explored populations. 

Each panel of Fig. \ref{fig:OFe} shows the distribution for the \hii\ regions located in galaxies of different morphological types. The observed distributions in each panel exhibit similar trends. In all cases, [O/Fe] decreases as [Fe/H] increases { following a linear trend with a slope $\sim$-0.7}. This trend is less defined, showing a broader dispersion, for the later morphological types (Sc and Sd). The described pattern is observed in all panels, despite the different number of \hii\ regions in each of them \citep[a consequence of the morphological distribution of the CALIFA galaxies, which peaks at Sb/Sbc, and the lower number of \hii\ regions in earlier morphological types e.g., ][]{lacerda20, espi20}. { Furthermore, the average [O/Fe] (top-left label of each panel in Fig. \ref{fig:OFe}), declines as galaxies have a later morphology. However, the [O/Fe] peak is not reached by the earlier morphological types (E/S0, [O/Fe]$\sim$0.23 dex), but for the Sa ones ([O/Fe]$\sim$0.27 dex). We should note that the difference between these two average values is small compared to their corresponding standard deviations ($\sim$0.26-0.21 dex), in any case. For later morphological types this value progressively declines with morphology.}

The decline of [O/Fe] as a function of [Fe/H] is qualitatively similar to the one for the stars in the Milky Way { \citep[MW, e.g.][]{yates12,grif21}}, not only for the oxygen abundance { \citep[e.g.][]{gaia}} but for any other $\alpha$-elements.
However, the quantities reported in those studies and the ones presented here have a different nature: abundance of individual stars (for the MW), and average properties of two different populations (in this study). To understand the nature of this decline and to demonstrate that we are tracing the same patterns we color-coded the distributions in Fig. 1 by the (\age) of the underlying stellar population. New patterns emerge from this exploration: ($i$) the decline of [O/Fe] as a function of [Fe/H] is much better defined for a fixed \age\ and morphology. { The nature of this relation will be discussed below}; and ($ii$) the broadening and shift  of the observed distribution { with morphology} is a consequence of the strong dependence of [O/Fe] with \age. {  Early-type (late-type) galaxies have older (younger) stellar populations in general with very little (strong) variation in the observed SFHs, and therefore a narrow (broad) range of \age\ values \citep[][]{ibarra16,rgb17}. This scenario fits with the observations for all morphological bins for early- and late-spirals (Sa to Sd/Sm). However, we find that E/S0 may present a slightly lower [O/Fe] than Sa galaxies (although the different may not be statistically significant).
If true, this could be due to a bias as our exploration is restricted to those galaxies with observed \hii\ region. For E/S0 galaxies they may be scarce \citep{espi20} and peculiar objects:
galaxies with either a SFH more extended than the bulk population of early-type ones, or suffering a rejuvenation by the capture of pristine gas \citep[e.g.][]{gomes16}. Nevertheless, despite of this possible bias, the result fits with the proposed connection between [O/Fe] and \age, since at least for the CALIFA sample the peak in ages for the stellar population is found for the Sa and not for the E/S0 galaxies \citep[e.g., Fig. 9 of][]{rgb17}.}

 
{ Similar trends between [$\alpha$/Fe], [Fe/H] and the age of the stellar populations} have been reported when exploring the $\alpha$-enhancement for resolved stars in the MW  { \citep[e.g.][]{apogee,lamost,galah,gaia}}, for { different compilations of early-type galaxies \citep{walcher15,walcher16},
and when comparing MW-like galaxies with simulations \citep[e.g.][]{calu09,yate12}}. In some cases they reported a weak decline in the metallicity with the stellar age \citep[e.g., Fig. 2][]{walcher16}, not appreciated in our results, due to a different definition of the parameters \citep[][]{coelho07}\footnote{in particular, [Z/H]=[Fe/H]+0.75[$\alpha$/Fe], instead of [Z/H]=[Fe/H]}. 
{ Finally, the stars in the disk of the MW presents a smoother relation of [O/Fe] with [Fe/H] than the one reported here \citep{galah}.}



{ We explore now if the observed} distributions are physically compatible with our current understanding of the chemical evolution in galaxies. To do so we made use of the ChEM by \citet{wein17}\footnote{\url{https://github.com/jobovy/kimmy}}, an analytic solution for the one-zone (fully mixed) model that incorporates a realistic delay time distribution for SNIa. In this way, it is possible to track the separate evolution of $\alpha$ and the iron peak elements. The model requires us
to define the shape of the SFH \citep[instead of the infall rate as other ChEMs do, e.g.][]{carigi19}, by selecting either an exponential ({\tt exp}, $e^{- t/\tau_*}$) or a linear-exponential ({\tt lin-exp}, $t e^{- t/\tau_*}$) functional form, both parametrized by the SF time delay ($\tau_*$). In addition, it requires us to define the depletion time $\tau_{\rm dep}$ and the mass-loading factor $\eta$, parameters that define the rate at which gas is transformed into stars and the amount of leaked  metals from the system as a function of the SFR \citep[e.g.][]{jkbb18}, respectively. This model has been successfully used to reproduce the distribution of \alf\ vs. [Fe/H] observed in the Milky-Way  \citep{wein17}.

{ Using this code, we generate a set of five ChEMs without attempting to fit them with the observations or to obtain a perfect one-to-one matching with the individual distributions reported for each morphological bin. We intend to reproduce the bulk observed distributions shown in Fig. \ref{fig:OFe} using a reasonable set of input parameters:} ($i$) { one model uses the the {\tt exp} functional-form to reproduce SFH of earlier-type galaxies (or the center of { early spirals}), and the remaining models a {\tt lin-exp} one that match better with the SFH of late-type galaxies} \citep[following][]{lopfer18,rgb18}; ($ii$) { a different value of $\tau_*$ is adopted for each of the five models spanning from 0.55 to 1.4 Gyr, to reproduce the range of observed of \age}; ($iii$) four different values for $\eta$ { were adopted}, ranging between 3 and 11, { covering the expected values for galaxies of different morphology and stellar masses, based in the relation between $\eta$ and M$_*$ by \citet{leet19}, and the morphology-mass relation in the CALIFA sample} \citep[e.g.][]{lacerda20}; finally, ($iv$) we adopt four different values of t$_{\rm dep}$, to reproduce the dependence of this parameter on the morphological type described by \citet{colombo18}, using the values tabulated by \citet{sanchez20}. {The top-inset of Fig. \ref{fig:OFe} indicates the parameters adopted for the five ChEMs.} For each model we derive the oxygen abundance of the last generation of stars, at each look-back time, which would correspond to our observed gas-phase [O/H], and the luminosity weighted [Fe/H] for the surviving stars formed before that time (i.e., older ages). To derive this latter parameter we are required to average the [Fe/H] weighted by the SFH, transforming the mass to light using the same M/L-ratio at each age adopted by {\sc Pipe3D}, and considering the mass-loss at each look-back time. The resulting tracks for each model, covering the last 2 Gyrs,
are represented in Fig. \ref{fig:OFe} as dot-dashed or solid lines (for each of the selected SFHs) and color-coded by the current \age. 

The simple ChEMs described before reproduce the main observed trends between [O/Fe] and [Fe/H].
In particular, ($i$) the average decline of the [O/Fe] as increasing [Fe/H], ($ii$) the range of [O/Fe] values, and ($iii$) its decline with the \age {, that induces the trends with morphology}, are well reproduced. It is worth noting that we did not fit/adjust the input parameters of the model to reproduce the observed abundance patterns: { $\eta$ and $\tau_{\rm dep}$ were extracted from the literature \citep{leet19,colombo18,sanchez20}, and the shape of the SFHs and $\tau_*$ were selected to reproduce just the {average \age}.} Based on this modelling, massive and earlier-type galaxies present stronger [O/Fe] enhancement, covering a narrower range of values, and with a better defined trend of this parameter with [Fe/H] as a consequence of a shorter and sharper SFH, with a low diversity in the SFH, { lower SFEs values} and having a better ability to retain metals. On the contrary, less-massive and later-type galaxies present weaker [O/Fe] enhancements for the opposite reasons: longer and steady SFH, a larger variety among them, { higher values of SFE} and a lower ability to retain metals. All these trends agree with our current understanding of galaxy evolution and the most recent results reported in the literature regarding the SFH, ChEH, MDFs and their relation with galaxy morphology, mass and location within galaxies \citep[e.g.][]{rgb17,rosa17,ibarra16,mejia20,camps20,ARAA}.

We should note that this is not the only interpretation possible for the observed distributions. { For instance,} using a more elaborated ChEM that assumes no outflows \citep{carigi19,Carigi2020}, it is possible to reproduce all the observed trends { by adopting a variable Initial Mass Function (IMF). The model assumes a certain gas infall rate as the basic regulator for the star formation. Changing this rate and adopting a \citet{salpeter55} IMF with different mass-ends (M$_{\rm cut}$=20-120 M$_\odot$), it is possible to reproduce the observed trends. The possibility of a variable IMF is a topic of current discussion \citep{kroupa01}, without a general consensus of what it is the main driver \citep[e.g.][]{navarro16}. Based on this model the observed trends would be the consequence of the ability of more massive galaxies to form more massive stars, higher M$_{\rm cut}$, than less massive ones, lower M$_{\rm cut}$ \citep{guna11,alvar08}.}


The comparison with these two particular ChEMs indicates that the observed trends (which are our primary results) are compatible with plausible scenarios/interpretations. Those scenarios are not unique, incompatible or exclusive. { Some of the observed trends 
were already predicted by the ChEMs presented in \citet{matt03}, as a consequence of the change in the SFH between early and late-type galaxies. However,} a change in the IMF does not exclude a possible change in the shape of the SFH or a different SFE. Finally, many of the parameters involved in those models are strongly degenerated, and modifying any of them may produce concordant results. However, these would not change the described observed trends, despite the detailed interpretation may be different.

\section{Conclusions}
\label{sec:con}

We present for the first time an exploration of the $\alpha$-enhancement in galaxies based on the comparison of the gas-phase oxygen abundance ([O/H]) with the stellar metallicity ([Z/H]) for a large sample of \hii\ regions and aggregations extracted for a representative sample of galaxies in the nearby Universe. From this exploration we show that:
   \begin{enumerate}
      \item\ [O/Fe] presents a decline with [Fe/H] similar to the one observed for \alf\ in the { MW, in early-type galaxies, and described in simulations}.
      \item The zero-point (slope)  of this relation, i.e., the absolute scale of [O/Fe], presents a strong (mild) dependence on both the stellar mass and morphology of the galaxy in agreement with { early scenarios, ChEMs, similations, and recent results.}
      \item We reproduce both trends using chemical evolution models by either (i) assuming that the SFH, SFE and $\eta$ or (ii) the  high-mass cut-off the IMF increases with the stellar mass of galaxies in agreement with { previous results.}
 \end{enumerate}

{ We have presented in this study the most relevant results of our on-going exploration. In a forthcoming article, Expinosa-Ponce et al. in prep, we will provide with further details on the modelling, exploring the reported trends galaxy by galaxy and in different regions within them.}  
 

\begin{acknowledgements}
{ We thank the referee for his/her comments and suggestions.}
SFS and J.B-B thanks CONACYT for grants CB-285080 and FC-2016-01-1916, and PAPIIT-DGAPA-IN100519 (UNAM) project. C.M. thanks UNAM/PAPIIT-IN101220. J.B-B thanks IA-100420 (DGAPA-PAPIIT ,UNAM) and CONACYT grant CF19-39578 support. L.G. thanks M.S.-Curie grant 839090. RGB acknowledges support from grants SEV-2017-0709 and P18-FRJ-2595. This study uses data provided by the Calar Alto Legacy
Integral Field Area (CALIFA) survey (http://califa.caha.es/), observed
at the Calar Alto Obsevatory.

\end{acknowledgements}

%
%
\bibliographystyle{aa}
\bibliography{my_bib} 

\begin{thebibliography}{67}
\expandafter\ifx\csname natexlab\endcsname\relax\def\natexlab#1{#1}\fi

\bibitem[{{Asplund} {et~al.}(2009){Asplund}, {Grevesse}, {Sauval}, \&
  {Scott}}]{aspl09}
{Asplund}, M., {Grevesse}, N., {Sauval}, A.~J., \& {Scott}, P. 2009, \araa, 47,
  481

\bibitem[{{Barrera-Ballesteros} {et~al.}(2018){Barrera-Ballesteros}, {Heckman},
  {S{\'a}nchez}, {Zakamska}, {Cleary}, {Zhu}, {Brinkmann}, {Drory}, \& {THE
  MaNGA TEAM}}]{jkbb18}
{Barrera-Ballesteros}, J.~K., {Heckman}, T., {S{\'a}nchez}, S.~F., {et~al.}
  2018, \apj, 852, 74

\bibitem[{{Calura} \& {Menci}(2009)}]{calu09}
{Calura}, F. \& {Menci}, N. 2009, \mnras, 400, 1347

\bibitem[{{Camps-Fari{\~n}a} {et~al.}(2021){Camps-Fari{\~n}a}, {Sanchez},
  {Lacerda}, {Carigi}, {Garc{\'\i}a-Benito}, {Mast}, \& {Galbany}}]{camps20}
{Camps-Fari{\~n}a}, A., {Sanchez}, S.~F., {Lacerda}, E.~A.~D., {et~al.} 2021,
  \mnras, 504, 3478

\bibitem[{{Carigi} {et~al.}(2020){Carigi}, {Peimbert}, {Peimbert}, \&
  {Delgado-Inglada}}]{Carigi2020}
{Carigi}, L., {Peimbert}, A., {Peimbert}, M., \& {Delgado-Inglada}, G. 2020,
  \rmxaa, 56, 235

\bibitem[{{Carigi} {et~al.}(2019){Carigi}, {Peimbert}, \&
  {Peimbert}}]{carigi19}
{Carigi}, L., {Peimbert}, M., \& {Peimbert}, A. 2019, \apj, 873, 107

\bibitem[{{Castrillo} {et~al.}(2021){Castrillo}, {Ascasibar}, {Galbany},
  {S{\'a}nchez}, {Badenes}, {Anderson}, {Kuncarayakti}, {Lyman}, \&
  {D{\'\i}az}}]{cast21}
{Castrillo}, A., {Ascasibar}, Y., {Galbany}, L., {et~al.} 2021, \mnras, 501,
  3122

\bibitem[{{Coelho} {et~al.}(2007){Coelho}, {Bruzual}, {Charlot}, {Weiss},
  {Barbuy}, \& {Ferguson}}]{coelho07}
{Coelho}, P., {Bruzual}, G., {Charlot}, S., {et~al.} 2007, \mnras, 382, 498

\bibitem[{{Colombo} {et~al.}(2018){Colombo}, {Kalinova}, {Utomo}, {Rosolowsky},
  {Bolatto}, {Levy}, {Wong}, {Sanchez}, {Leroy}, {Ostriker}, {Blitz}, {Vogel},
  {Mast}, {Garc{\'{\i}}a-Benito}, {Husemann}, {Dannerbauer}, {Ellmeier}, \&
  {Cao}}]{colombo18}
{Colombo}, D., {Kalinova}, V., {Utomo}, D., {et~al.} 2018, \mnras, 475, 1791

\bibitem[{{Comte}(1975)}]{comte74}
{Comte}, G. 1975, \aap, 39, 197

\bibitem[{{Conroy} \& {van Dokkum}(2012)}]{conroy12}
{Conroy}, C. \& {van Dokkum}, P.~G. 2012, \apj, 760, 71

\bibitem[{{Espinosa-Ponce} {et~al.}(2020){Espinosa-Ponce}, {S{\'a}nchez},
  {Morisset}, {Barrera-Ballesteros}, {Galbany}, {Garc{\'\i}a-Benito},
  {Lacerda}, \& {Mast}}]{espi20}
{Espinosa-Ponce}, C., {S{\'a}nchez}, S.~F., {Morisset}, C., {et~al.} 2020,
  \mnras, 494, 1622

\bibitem[{{Esteban} \& {Garc{\'\i}a-Rojas}(2018)}]{esteban18}
{Esteban}, C. \& {Garc{\'\i}a-Rojas}, J. 2018, \mnras, 478, 2315

\bibitem[{{Fern{\'a}ndez-Alvar} {et~al.}(2018){Fern{\'a}ndez-Alvar}, {Carigi},
  {Schuster}, {Hayes}, {{\'A}vila-Vergara}, {Majewski}, {Allende Prieto},
  {Beers}, {S{\'a}nchez}, {Zamora}, {Garc{\'\i}a-Hern{\'a}ndez}, {Tang},
  {Fern{\'a}ndez-Trincado}, {Tissera}, {Geisler}, \& {Villanova}}]{alvar08}
{Fern{\'a}ndez-Alvar}, E., {Carigi}, L., {Schuster}, W.~J., {et~al.} 2018,
  \apj, 852, 50

\bibitem[{{Franchini} {et~al.}(2021){Franchini}, {Morossi}, {Di Marcantonio},
  {Chavez}, {Adibekyan}, {Bensby}, {Bragaglia}, {Gonneau}, {Heiter},
  {Kordopatis}, {Magrini}, {Romano}, {Sbordone}, {Smiljanic},
  {Tautvai{\v{s}}ien{\.{e}}}, {Gilmore}, {Randich}, {Bayo}, {Carraro},
  {Morbidelli}, \& {Zaggia}}]{gaia}
{Franchini}, M., {Morossi}, C., {Di Marcantonio}, P., {et~al.} 2021, \aj, 161,
  9

\bibitem[{{Galbany} {et~al.}(2018){Galbany}, {Anderson}, {S{\'a}nchez},
  {Kuncarayakti}, {Pedraz}, {Gonz{\'a}lez-Gait{\'a}n}, {Stanishev},
  {Dom{\'\i}nguez}, {Moreno-Raya}, {Wood-Vasey}, {Mour{\~a}o}, {Ponder},
  {Badenes}, {Moll{\'a}}, {L{\'o}pez-S{\'a}nchez}, {Rosales-Ortega},
  {V{\'\i}lchez}, {Garc{\'\i}a-Benito}, \& {Marino}}]{pisco}
{Galbany}, L., {Anderson}, J.~P., {S{\'a}nchez}, S.~F., {et~al.} 2018, \apj,
  855, 107

\bibitem[{{Galbany} {et~al.}(2014){Galbany}, {Stanishev}, {Mour{\~a}o},
  {Rodrigues}, {Flores}, {Garc{\'{\i}}a-Benito}, {Mast}, {Mendoza},
  {S{\'a}nchez}, {Badenes}, {Barrera-Ballesteros}, {Bland-Hawthorn},
  {Falc{\'o}n-Barroso}, {Garc{\'{\i}}a-Lorenzo}, {Gomes}, {Gonz{\'a}lez
  Delgado}, {Kehrig}, {Lyubenova}, {L{\'o}pez-S{\'a}nchez}, {de
  Lorenzo-C{\'a}ceres}, {Marino}, {Meidt}, {Moll{\'a}}, {Papaderos},
  {P{\'e}rez-Torres}, {Rosales-Ortega}, \& {van de Ven}}]{galbany:2014}
{Galbany}, L., {Stanishev}, V., {Mour{\~a}o}, A.~M., {et~al.} 2014, \aap, 572,
  A38

\bibitem[{{Garc{\'{\i}}a-Benito} {et~al.}(2017){Garc{\'{\i}}a-Benito},
  {Gonz{\'a}lez Delgado}, {P{\'e}rez}, {Cid Fernandes}, {Cortijo-Ferrero},
  {L{\'o}pez Fern{\'a}ndez}, {de Amorim}, {Lacerda}, {Vale Asari}, \&
  {S{\'a}nchez}}]{rgb17}
{Garc{\'{\i}}a-Benito}, R., {Gonz{\'a}lez Delgado}, R.~M., {P{\'e}rez}, E.,
  {et~al.} 2017, \aap, 608, A27

\bibitem[{{Garc{\'\i}a-Benito} {et~al.}(2019){Garc{\'\i}a-Benito},
  {Gonz{\'a}lez Delgado}, {P{\'e}rez}, {Cid Fernandes}, {S{\'a}nchez}, \& {de
  Amorim}}]{rgb18}
{Garc{\'\i}a-Benito}, R., {Gonz{\'a}lez Delgado}, R.~M., {P{\'e}rez}, E.,
  {et~al.} 2019, \aap, 621, A120

\bibitem[{{Gomes} {et~al.}(2016){Gomes}, {Papaderos}, {V{\'{\i}}lchez},
  {Kehrig}, {Iglesias-P{\'a}ramo}, {Breda}, {Lehnert}, {S{\'a}nchez},
  {Ziegler}, {Dos Reis}, {Bland-Hawthorn}, {Galbany}, {Bomans},
  {Rosales-Ortega}, {Walcher}, {Garc{\'{\i}}a-Benito}, {M{\'a}rquez}, {Del
  Olmo}, {Moll{\'a}}, {Marino}, {Catal{\'a}n-Torrecilla}, {Gonz{\'a}lez
  Delgado}, {L{\'o}pez-S{\'a}nchez}, \& {Califa Collaboration}}]{gomes16}
{Gomes}, J.~M., {Papaderos}, P., {V{\'{\i}}lchez}, J.~M., {et~al.} 2016, \aap,
  585, A92

\bibitem[{{Gonz{\'a}lez Delgado} {et~al.}(2014){Gonz{\'a}lez Delgado}, {Cid
  Fernandes}, {Garc{\'{\i}}a-Benito}, {P{\'e}rez}, {de Amorim},
  {Cortijo-Ferrero}, {Lacerda}, {L{\'o}pez Fern{\'a}ndez}, {S{\'a}nchez}, {Vale
  Asari}, {Alves}, {Bland-Hawthorn}, {Galbany}, {Gallazzi}, {Husemann},
  {Bekeraite}, {Jungwiert}, {L{\'o}pez-S{\'a}nchez}, {de Lorenzo-C{\'a}ceres},
  {Marino}, {Mast}, {Moll{\'a}}, {del Olmo}, {S{\'a}nchez-Bl{\'a}zquez}, {van
  de Ven}, {V{\'{\i}}lchez}, {Walcher}, {Wisotzki}, {Ziegler}, \&
  {collaboration920}}]{rosa14b}
{Gonz{\'a}lez Delgado}, R.~M., {Cid Fernandes}, R., {Garc{\'{\i}}a-Benito}, R.,
  {et~al.} 2014, \apjl, 791, L16

\bibitem[{{Gonz{\'a}lez Delgado} {et~al.}(2017){Gonz{\'a}lez Delgado},
  {P{\'e}rez}, {Cid Fernandes}, {Garc{\'{\i}}a-Benito}, {L{\'o}pez
  Fern{\'a}ndez}, {Vale Asari}, {Cortijo-Ferrero}, {de Amorim}, {Lacerda},
  {S{\'a}nchez}, {Lehnert}, \& {Walcher}}]{rosa17}
{Gonz{\'a}lez Delgado}, R.~M., {P{\'e}rez}, E., {Cid Fernandes}, R., {et~al.}
  2017, \aap, 607, A128

\bibitem[{{Griffith} {et~al.}(2021){Griffith}, {Weinberg}, {Johnson}, {Beaton},
  {Garc{\'\i}a-Hern{\'a}ndez}, {Hasselquist}, {Holtzman}, {Johnson},
  {J{\"o}nsson}, {Lane}, {Nataf}, \& {Roman-Lopes}}]{grif21}
{Griffith}, E., {Weinberg}, D.~H., {Johnson}, J.~A., {et~al.} 2021, \apj, 909,
  77

\bibitem[{{Gunawardhana} {et~al.}(2011){Gunawardhana}, {Hopkins}, {Sharp},
  {Brough}, {Taylor}, {Bland-Hawthorn}, {Maraston}, {Tuffs}, {Popescu},
  {Wijesinghe}, {Jones}, {Croom}, {Sadler}, {Wilkins}, {Driver}, {Liske},
  {Norberg}, {Baldry}, {Bamford}, {Loveday}, {Peacock}, {Robotham}, {Zucker},
  {Parker}, {Conselice}, {Cameron}, {Frenk}, {Hill}, {Kelvin}, {Kuijken},
  {Madore}, {Nichol}, {Parkinson}, {Pimbblet}, {Prescott}, {Sutherland},
  {Thomas}, \& {van Kampen}}]{guna11}
{Gunawardhana}, M.~L.~P., {Hopkins}, A.~M., {Sharp}, R.~G., {et~al.} 2011,
  \mnras, 415, 1647

\bibitem[{{Hayden} {et~al.}(2015){Hayden}, {Bovy}, {Holtzman}, {Nidever},
  {Bird}, {Weinberg}, {Andrews}, {Majewski}, {Allende Prieto}, {Anders},
  {Beers}, {Bizyaev}, {Chiappini}, {Cunha}, {Frinchaboy},
  {Garc{\'\i}a-Her{\'n}andez}, {Garc{\'\i}a P{\'e}rez}, {Girardi}, {Harding},
  {Hearty}, {Johnson}, {M{\'e}sz{\'a}ros}, {Minchev}, {O'Connell}, {Pan},
  {Robin}, {Schiavon}, {Schneider}, {Schultheis}, {Shetrone}, {Skrutskie},
  {Steinmetz}, {Smith}, {Wilson}, {Zamora}, \& {Zasowski}}]{apogee}
{Hayden}, M.~R., {Bovy}, J., {Holtzman}, J.~A., {et~al.} 2015, \apj, 808, 132

\bibitem[{{Ho}(2019)}]{ho19}
{Ho}, I.~T. 2019, \mnras, 485, 3569

\bibitem[{{Ibarra-Medel} {et~al.}(2019){Ibarra-Medel}, {Avila-Reese},
  {S{\'a}nchez}, {Gonz{\'a}lez-Samaniego}, \&
  {Rodr{\'{\i}}guez-Puebla}}]{ibarra19}
{Ibarra-Medel}, H.~J., {Avila-Reese}, V., {S{\'a}nchez}, S.~F.,
  {Gonz{\'a}lez-Samaniego}, A., \& {Rodr{\'{\i}}guez-Puebla}, A. 2019, \mnras,
  483, 4525

\bibitem[{{Ibarra-Medel} {et~al.}(2016){Ibarra-Medel}, {S{\'a}nchez},
  {Avila-Reese}, {Hern{\'a}ndez-Toledo}, {Gonz{\'a}lez}, {Drory}, {Bundy},
  {Bizyaev}, {Cano-D{\'{\i}}az}, {Malanushenko}, {Pan}, {Roman-Lopes}, \&
  {Thomas}}]{ibarra16}
{Ibarra-Medel}, H.~J., {S{\'a}nchez}, S.~F., {Avila-Reese}, V., {et~al.} 2016,
  \mnras, 463, 2799

\bibitem[{{Kelz} {et~al.}(2006){Kelz}, {Verheijen}, {Roth}, {Bauer}, {Becker},
  {Paschke}, {Popow}, {S{\'a}nchez}, \& {Laux}}]{kelz06}
{Kelz}, A., {Verheijen}, M.~A.~W., {Roth}, M.~M., {et~al.} 2006, \pasp, 118,
  129

\bibitem[{{Kobayashi} {et~al.}(2020){Kobayashi}, {Karakas}, \&
  {Lugaro}}]{koba20}
{Kobayashi}, C., {Karakas}, A.~I., \& {Lugaro}, M. 2020, \apj, 900, 179

\bibitem[{{Kroupa}(2001)}]{kroupa01}
{Kroupa}, P. 2001, \mnras, 322, 231

\bibitem[{{Lacerda} {et~al.}(2020){Lacerda}, {S{\'a}nchez}, {Cid Fernandes},
  {L{\'o}pez-Cob{\'a}}, {Espinosa-Ponce}, \& {Galbany}}]{lacerda20}
{Lacerda}, E. A.~D., {S{\'a}nchez}, S.~F., {Cid Fernandes}, R., {et~al.} 2020,
  \mnras, 492, 3073

\bibitem[{{Leethochawalit} {et~al.}(2019){Leethochawalit}, {Kirby}, {Ellis},
  {Moran}, \& {Treu}}]{leet19}
{Leethochawalit}, N., {Kirby}, E.~N., {Ellis}, R.~S., {Moran}, S.~M., \&
  {Treu}, T. 2019, \apj, 885, 100

\bibitem[{{Lian} {et~al.}(2018){Lian}, {Thomas}, {Maraston}, {Goddard},
  {Parikh}, {Fern{\'a}ndez-Trincado}, {Roman-Lopes}, {Rong}, {Tang}, \&
  {Yan}}]{lian18}
{Lian}, J., {Thomas}, D., {Maraston}, C., {et~al.} 2018, \mnras, 476, 3883

\bibitem[{{L{\'o}pez Fern{\'a}ndez} {et~al.}(2018){L{\'o}pez Fern{\'a}ndez},
  {Gonz{\'a}lez Delgado}, {P{\'e}rez}, {Garc{\'{\i}}a-Benito}, {Cid Fernandes},
  {Schoenell}, {S{\'a}nchez}, {Gallazzi}, {S{\'a}nchez-Bl{\'a}zquez}, {Vale
  Asari}, \& {Walcher}}]{lopfer18}
{L{\'o}pez Fern{\'a}ndez}, R., {Gonz{\'a}lez Delgado}, R.~M., {P{\'e}rez}, E.,
  {et~al.} 2018, \aap, 615, A27

\bibitem[{{Maiolino} \& {Mannucci}(2019)}]{maio19}
{Maiolino}, R. \& {Mannucci}, F. 2019, \aapr, 27, 3

\bibitem[{{Mart{\'{\i}}n-Navarro} {et~al.}(2015){Mart{\'{\i}}n-Navarro},
  {Vazdekis}, {La Barbera}, {Falc{\'o}n-Barroso}, {Lyubenova}, {van de Ven},
  {Ferreras}, {S{\'a}nchez}, {Trager}, {Garc{\'{\i}}a-Benito}, {Mast},
  {Mendoza}, {S{\'a}nchez-Bl{\'a}zquez}, {Gonz{\'a}lez Delgado}, {Walcher}, \&
  {The CALIFA Team}}]{navarro16}
{Mart{\'{\i}}n-Navarro}, I., {Vazdekis}, A., {La Barbera}, F., {et~al.} 2015,
  \apjl, 806, L31

\bibitem[{{Matteucci}(1992)}]{matt92}
{Matteucci}, F. 1992, \memsai, 63, 301

\bibitem[{{Matteucci}(2003)}]{matt03}
{Matteucci}, F. 2003, \apss, 284, 539

\bibitem[{{Matteucci} {et~al.}(1999){Matteucci}, {Romano}, \&
  {Molaro}}]{matt99}
{Matteucci}, F., {Romano}, D., \& {Molaro}, P. 1999, \aap, 341, 458

\bibitem[{{Mej{\'\i}a-Narv{\'a}ez} {et~al.}(2020){Mej{\'\i}a-Narv{\'a}ez},
  {S{\'a}nchez}, {Lacerda}, {Carigi}, {Galbany}, {Husemann}, \&
  {Garc{\'\i}a-Benito}}]{mejia20}
{Mej{\'\i}a-Narv{\'a}ez}, A., {S{\'a}nchez}, S.~F., {Lacerda}, E.~A.~D.,
  {et~al.} 2020, \mnras, 499, 4838

\bibitem[{{Nandakumar} {et~al.}(2020){Nandakumar}, {Hayden}, {Sharma}, {Buder},
  {Asplund}, {Bland-Hawthorn}, {De Silva}, {D'Orazi}, {Freeman}, {Kos},
  {Lewis}, {Martell}, {Schlesinger}, {Lin}, {Simpson}, {Zucker}, {Zwitter},
  {Nordlander}, {Casagrande}, {Lind}, {Cotar}, {Stello}, {Wittenmyer}, \&
  {Tepper-Garcia}}]{galah}
{Nandakumar}, G., {Hayden}, M.~R., {Sharma}, S., {et~al.} 2020, arXiv e-prints,
  arXiv:2011.02783

\bibitem[{{Peimbert} \& {Peimbert}(2010)}]{peim10}
{Peimbert}, A. \& {Peimbert}, M. 2010, \apj, 724, 791

\bibitem[{{Peimbert} {et~al.}(1978){Peimbert}, {Torres-Peimbert}, \&
  {Rayo}}]{peim78}
{Peimbert}, M., {Torres-Peimbert}, S., \& {Rayo}, J.~F. 1978, \apj, 220, 516

\bibitem[{{Salpeter}(1955)}]{salpeter55}
{Salpeter}, E.~E. 1955, \apj, 121, 161

\bibitem[{{S{\'a}nchez}(2020)}]{ARAA}
{S{\'a}nchez}, S.~F. 2020, \araa, 58, 99

\bibitem[{{S{\'a}nchez} {et~al.}(2016{\natexlab{a}}){S{\'a}nchez},
  {Garc{\'{\i}}a-Benito}, {Zibetti}, {Walcher}, {Husemann}, {Mendoza},
  {Galbany}, {Falc{\'o}n-Barroso}, {Mast}, {Aceituno}, {Aguerri}, {Alves},
  {Amorim}, {Ascasibar}, {Barrado-Navascues}, {Barrera-Ballesteros},
  {Bekerait{\`e}}, {Bland-Hawthorn}, {Cano D{\'{\i}}az}, {Cid Fernandes},
  {Cavichia}, {Cortijo}, {Dannerbauer}, {Demleitner}, {D{\'{\i}}az}, {Dettmar},
  {de Lorenzo-C{\'a}ceres}, {del Olmo}, {Galazzi}, {Garc{\'{\i}}a-Lorenzo},
  {Gil de Paz}, {Gonz{\'a}lez Delgado}, {Holmes}, {Igl{\'e}sias-P{\'a}ramo},
  {Kehrig}, {Kelz}, {Kennicutt}, {Kleemann}, {Lacerda}, {L{\'o}pez
  Fern{\'a}ndez}, {L{\'o}pez S{\'a}nchez}, {Lyubenova}, {Marino},
  {M{\'a}rquez}, {Mendez-Abreu}, {Moll{\'a}}, {Monreal-Ibero}, {Ortega
  Minakata}, {Torres-Papaqui}, {P{\'e}rez}, {Rosales-Ortega}, {Roth},
  {S{\'a}nchez-Bl{\'a}zquez}, {Schilling}, {Spekkens}, {Vale Asari}, {van den
  Bosch}, {van de Ven}, {Vilchez}, {Wild}, {Wisotzki}, {Y{\i}ld{\i}r{\i}m}, \&
  {Ziegler}}]{sanchez16}
{S{\'a}nchez}, S.~F., {Garc{\'{\i}}a-Benito}, R., {Zibetti}, S., {et~al.}
  2016{\natexlab{a}}, \aap, 594, A36

\bibitem[{{S{\'a}nchez} {et~al.}(2012{\natexlab{a}}){S{\'a}nchez}, {Kennicutt},
  {Gil de Paz}, {van de Ven}, {V{\'{\i}}lchez}, {Wisotzki}, {Walcher}, {Mast},
  {Aguerri}, {Albiol-P{\'e}rez}, {Alonso-Herrero}, {Alves}, {Bakos},
  {Bart{\'a}kov{\'a}}, {Bland-Hawthorn}, {Boselli}, {Bomans},
  {Castillo-Morales}, {Cortijo-Ferrero}, {de Lorenzo-C{\'a}ceres}, {Del Olmo},
  {Dettmar}, {D{\'{\i}}az}, {Ellis}, {Falc{\'o}n-Barroso}, {Flores},
  {Gallazzi}, {Garc{\'{\i}}a-Lorenzo}, {Gonz{\'a}lez Delgado}, {Gruel},
  {Haines}, {Hao}, {Husemann}, {Igl{\'e}sias-P{\'a}ramo}, {Jahnke}, {Johnson},
  {Jungwiert}, {Kalinova}, {Kehrig}, {Kupko}, {L{\'o}pez-S{\'a}nchez},
  {Lyubenova}, {Marino}, {M{\'a}rmol-Queralt{\'o}}, {M{\'a}rquez}, {Masegosa},
  {Meidt}, {Mendez-Abreu}, {Monreal-Ibero}, {Montijo}, {Mour{\~a}o},
  {Palacios-Navarro}, {Papaderos}, {Pasquali}, {Peletier}, {P{\'e}rez},
  {P{\'e}rez}, {Quirrenbach}, {Rela{\~n}o}, {Rosales-Ortega}, {Roth},
  {Ruiz-Lara}, {S{\'a}nchez-Bl{\'a}zquez}, {Sengupta}, {Singh}, {Stanishev},
  {Trager}, {Vazdekis}, {Viironen}, {Wild}, {Zibetti}, \& {Ziegler}}]{califa}
{S{\'a}nchez}, S.~F., {Kennicutt}, R.~C., {Gil de Paz}, A., {et~al.}
  2012{\natexlab{a}}, \aap, 538, A8

\bibitem[{{S{\'a}nchez} {et~al.}(2015){S{\'a}nchez}, {P{\'e}rez},
  {Rosales-Ortega}, {Miralles-Caballero}, {L{\'o}pez-S{\'a}nchez},
  {Iglesias-P{\'a}ramo}, {Marino}, {S{\'a}nchez-Menguiano},
  {Garc{\'{\i}}a-Benito}, {Mast}, {Mendoza}, {Papaderos}, {Ellis}, {Galbany},
  {Kehrig}, {Monreal-Ibero}, {Gonz{\'a}lez Delgado}, {Moll{\'a}}, {Ziegler},
  {de Lorenzo-C{\'a}ceres}, {Mendez-Abreu}, {Bland-Hawthorn}, {Bekerait{\.e}},
  {Roth}, {Pasquali}, {D{\'{\i}}az}, {Bomans}, {van de Ven}, \&
  {Wisotzki}}]{sanchez15}
{S{\'a}nchez}, S.~F., {P{\'e}rez}, E., {Rosales-Ortega}, F.~F., {et~al.} 2015,
  \aap, 574, A47

\bibitem[{{S{\'a}nchez} {et~al.}(2016{\natexlab{b}}){S{\'a}nchez}, {P{\'e}rez},
  {S{\'a}nchez-Bl{\'a}zquez}, {Garc{\'{\i}}a-Benito}, {Ibarra-Mede},
  {Gonz{\'a}lez}, {Rosales-Ortega}, {S{\'a}nchez-Menguiano}, {Ascasibar},
  {Bitsakis}, {Law}, {Cano-D{\'{\i}}az}, {L{\'o}pez-Cob{\'a}}, {Marino}, {Gil
  de Paz}, {L{\'o}pez-S{\'a}nchez}, {Barrera-Ballesteros}, {Galbany}, {Mast},
  {Abril-Melgarejo}, \& {Roman-Lopes}}]{pipe3d_ii}
{S{\'a}nchez}, S.~F., {P{\'e}rez}, E., {S{\'a}nchez-Bl{\'a}zquez}, P., {et~al.}
  2016{\natexlab{b}}, \rmxaa, 52, 171

\bibitem[{{S{\'a}nchez} {et~al.}(2014){S{\'a}nchez}, {Rosales-Ortega},
  {Iglesias-P{\'a}ramo}, {Moll{\'a}}, {Barrera-Ballesteros}, {Marino},
  {P{\'e}rez}, {S{\'a}nchez-Blazquez}, {Gonz{\'a}lez Delgado}, {Cid Fernandes},
  {de Lorenzo-C{\'a}ceres}, {Mendez-Abreu}, {Galbany}, {Falcon-Barroso},
  {Miralles-Caballero}, {Husemann}, {Garc{\'{\i}}a-Benito}, {Mast}, {Walcher},
  {Gil de Paz}, {Garc{\'{\i}}a-Lorenzo}, {Jungwiert}, {V{\'{\i}}lchez},
  {J{\'{\i}}lkov{\'a}}, {Lyubenova}, {Cortijo-Ferrero}, {D{\'{\i}}az},
  {Wisotzki}, {M{\'a}rquez}, {Bland-Hawthorn}, {Ellis}, {van de Ven}, {Jahnke},
  {Papaderos}, {Gomes}, {Mendoza}, \& {L{\'o}pez-S{\'a}nchez}}]{sanchez14}
{S{\'a}nchez}, S.~F., {Rosales-Ortega}, F.~F., {Iglesias-P{\'a}ramo}, J.,
  {et~al.} 2014, \aap, 563, A49

\bibitem[{{S{\'a}nchez} {et~al.}(2012{\natexlab{b}}){S{\'a}nchez},
  {Rosales-Ortega}, {Marino}, {Iglesias-P{\'a}ramo}, {V{\'{\i}}lchez},
  {Kennicutt}, {D{\'{\i}}az}, {Mast}, {Monreal-Ibero}, {Garc{\'{\i}}a-Benito},
  {Bland-Hawthorn}, {P{\'e}rez}, {Gonz{\'a}lez Delgado}, {Husemann},
  {L{\'o}pez-S{\'a}nchez}, {Cid Fernandes}, {Kehrig}, {Walcher}, {Gil de Paz},
  \& {Ellis}}]{sanchez12b}
{S{\'a}nchez}, S.~F., {Rosales-Ortega}, F.~F., {Marino}, R.~A., {et~al.}
  2012{\natexlab{b}}, \aap, 546, A2

\bibitem[{{S{\'a}nchez} {et~al.}(2021){S{\'a}nchez}, {Walcher},
  {Lopez-Cob{\'a}}, {Barrera-Ballesteros}, {Mej{\'\i}a-Narv{\'a}ez},
  {Espinosa-Ponce}, \& {Camps-Fari{\~n}a}}]{sanchez20}
{S{\'a}nchez}, S.~F., {Walcher}, C.~J., {Lopez-Cob{\'a}}, C., {et~al.} 2021,
  \rmxaa, 57, 3

\bibitem[{{Searle}(1971)}]{sear71}
{Searle}, L. 1971, \apj, 168, 327

\bibitem[{{Thomas} {et~al.}(2005){Thomas}, {Maraston}, {Bender}, \& {Mendes de
  Oliveira}}]{thomas05}
{Thomas}, D., {Maraston}, C., {Bender}, R., \& {Mendes de Oliveira}, C. 2005,
  \apj, 621, 673

\bibitem[{{Trager} {et~al.}(2000){Trager}, {Faber}, {Worthey}, \&
  {Gonz{\'a}lez}}]{trager+00}
{Trager}, S.~C., {Faber}, S.~M., {Worthey}, G., \& {Gonz{\'a}lez}, J.~J. 2000,
  \aj, 119, 1645

\bibitem[{{Vazdekis} {et~al.}(2015){Vazdekis}, {Coelho}, {Cassisi},
  {Ricciardelli}, {Falc{\'o}n-Barroso}, {S{\'a}nchez-Bl{\'a}zquez}, {Barbera},
  {Beasley}, \& {Pietrinferni}}]{vazdekis15}
{Vazdekis}, A., {Coelho}, P., {Cassisi}, S., {et~al.} 2015, \mnras, 449, 1177

\bibitem[{{Walcher} {et~al.}(2009){Walcher}, {Coelho}, {Gallazzi}, \&
  {Charlot}}]{walcher09}
{Walcher}, C.~J., {Coelho}, P., {Gallazzi}, A., \& {Charlot}, S. 2009, \mnras,
  398, L44

\bibitem[{{Walcher} {et~al.}(2015){Walcher}, {Coelho}, {Gallazzi}, {Bruzual},
  {Charlot}, \& {Chiappini}}]{walcher15}
{Walcher}, C.~J., {Coelho}, P.~R.~T., {Gallazzi}, A., {et~al.} 2015, \aap, 582,
  A46

\bibitem[{{Walcher} {et~al.}(2014){Walcher}, {Wisotzki}, {Bekerait{\'e}},
  {Husemann}, {Iglesias-P{\'a}ramo}, {Backsmann}, {Barrera Ballesteros},
  {Catal{\'a}n-Torrecilla}, {Cortijo}, {del Olmo}, {Garcia Lorenzo},
  {Falc{\'o}n-Barroso}, {Jilkova}, {Kalinova}, {Mast}, {Marino},
  {M{\'e}ndez-Abreu}, {Pasquali}, {S{\'a}nchez}, {Trager}, {Zibetti},
  {Aguerri}, {Alves}, {Bland-Hawthorn}, {Boselli}, {Castillo Morales}, {Cid
  Fernandes}, {Flores}, {Galbany}, {Gallazzi}, {Garc{\'{\i}}a-Benito}, {Gil de
  Paz}, {Gonz{\'a}lez-Delgado}, {Jahnke}, {Jungwiert}, {Kehrig}, {Lyubenova},
  {M{\'a}rquez Perez}, {Masegosa}, {Monreal Ibero}, {P{\'e}rez}, {Quirrenbach},
  {Rosales-Ortega}, {Roth}, {Sanchez-Blazquez}, {Spekkens}, {Tundo}, {van de
  Ven}, {Verheijen}, {Vilchez}, \& {Ziegler}}]{walcher14}
{Walcher}, C.~J., {Wisotzki}, L., {Bekerait{\'e}}, S., {et~al.} 2014, \aap,
  569, A1

\bibitem[{{Walcher} {et~al.}(2016){Walcher}, {Yates}, {Minchev}, {Chiappini},
  {Bergemann}, {Bruzual}, {Charlot}, {Coelho}, {Gallazzi}, \&
  {Martig}}]{walcher16}
{Walcher}, C.~J., {Yates}, R.~M., {Minchev}, I., {et~al.} 2016, \aap, 594, A61

\bibitem[{{Watson} {et~al.}(2021){Watson}, {Davies}, {Brough}, {Croom},
  {D'Eugenio}, {Glazebrook}, {Groves}, {L{\'o}pez-S{\'a}nchez}, {van de Sande},
  {Scott}, {Vaughan}, {Walcher}, {Bland-Hawthorn}, {Bryant}, {Goodwin},
  {Lawrence}, {Lorente}, {Owers}, \& {Richards}}]{watson21}
{Watson}, P.~J., {Davies}, R.~L., {Brough}, S., {et~al.} 2021, arXiv e-prints,
  arXiv:2106.01928

\bibitem[{{Weinberg} {et~al.}(2017){Weinberg}, {Andrews}, \&
  {Freudenburg}}]{wein17}
{Weinberg}, D.~H., {Andrews}, B.~H., \& {Freudenburg}, J. 2017, \apj, 837, 183

\bibitem[{{Woosley} \& {Weaver}(1995)}]{woos95}
{Woosley}, S.~E. \& {Weaver}, T.~A. 1995, \apjs, 101, 181

\bibitem[{{Yates} {et~al.}(2013){Yates}, {Henriques}, {Thomas}, {Kauffmann},
  {Johansson}, \& {White}}]{yates12}
{Yates}, R.~M., {Henriques}, B., {Thomas}, P.~A., {et~al.} 2013, \mnras, 435,
  3500

\bibitem[{{Yates} {et~al.}(2012){Yates}, {Kauffmann}, \& {Guo}}]{yate12}
{Yates}, R.~M., {Kauffmann}, G., \& {Guo}, Q. 2012, \mnras, 422, 215

\bibitem[{{Yu} {et~al.}(2021){Yu}, {Li}, {Chen}, {Huang}, {Jia}, {Xiang},
  {Yuan}, {Shi}, {Wang}, \& {Liu}}]{lamost}
{Yu}, Z., {Li}, J., {Chen}, B., {et~al.} 2021, \apj, 912, 106

\end{thebibliography}

%

\end{document}